%% file: main.tex
\documentclass[journal]{IEEEtran}

\usepackage{subfiles}

\usepackage[dvipdfmx]{graphicx}
\usepackage{latexsym}
\usepackage{url}

\begin{document}

\title{Model Extraction Attacks \\ against Recurrent Neural Networks}



\author{Tatsuya Takemura, Naoto Yanai and Toru Fujiwara        
\thanks{Tatsuya Takemura, Naoto Yanai and Toru Fujiwara are with Graduate School of Information Science and Technology, Osaka University e-mail: (see http://www-infosec.ist.osaka-u.ac.jp/member.html).}
\thanks{}
}


\maketitle

\begin{abstract}
Model extraction attacks are a kind of attacks in which an adversary obtains a new model, whose performance is equivalent to that of a target model, via query access to the target model \textit{efficiently}, i.e., fewer datasets and computational resources than those of the target model. Existing works have dealt with only simple deep neural networks (DNNs), e.g., only three layers, as targets of model extraction attacks, and hence are not aware of the effectiveness of recurrent neural networks (RNNs) in dealing with time-series data. In this work, we shed light on the threats of model extraction attacks against RNNs. We discuss whether a model with a higher accuracy can be extracted with a simple RNN from a long short-term memory (LSTM), which is a more complicated and powerful RNN. Specifically, we tackle the following problems. First, in a case of a classification problem, such as image recognition, extraction of an RNN model without final outputs from an LSTM model is presented by utilizing outputs halfway through the sequence. Next, in a case of a regression problem. such as in weather forecasting, a new attack by newly configuring a loss function is presented. We conduct experiments on our model extraction attacks against an RNN and an LSTM trained with publicly available academic datasets. We then show that a model with a higher accuracy can be extracted efficiently, especially through configuring a loss function and a more complex architecture different from the target model. 
\end{abstract}


\IEEEpeerreviewmaketitle

\subfile{sec1}

\subfile{sec2}

\subfile{sec3}

\subfile{sec4}

\subfile{sec5}

\subfile{sec6}

\section*{Acknowledgment}
This work is supported by the Cabinet Office (CAO), Cross-ministerial
Strategic Innovation Promotion Program (SIP), Cyber Physical Security
for IoT Society (funding agency: NEDO). 
We also would like to thank Jason Paul Cruz for helpful comments.

\bibliographystyle{unsrt}
\bibliography{main} 

\end{document}

%% file: sec1.tex
\section{Introduction}
\subsection{Backgrounds}
Deep learning is a state-of-the-art technology for machine learning and is known to provide various advantages in many areas. 
Deep learning requires heavy computations, and thus a business style called machine-learning-as-a-service (MLaaS), where a machine learning model is hosted via a public server, has received attention recently. Well-know MLaaS include AWS\footnote{\url{https://aws.amazon.com/jp/aml/}} and 
Microsoft Azure\footnote{\url{https://azure.microsoft.com/ja-jp/services/machine-
learning-studio/}}. In such a situation where a machine learning model consists of two tasks, i.e., \textit{training} and \textit{prediction}, a trained model is stored in a public server, e.g., cloud server, and a client requests the model to run a prediction task via APIs. 

However, the execution of prediction tasks via APIs may leak information about a model to a client. \textit{Model extraction attacks}~\cite{FFA16} have received attention in recent years from the standpoint of information leakage described above. 
In particular, an adversary who behaves as a client trains his/her own model by utilizing APIs of a machine learning model hosted by a public server, called an \textit{original model}, and its prediction results. The trained model by the adversary is called a \textit{substitute model}. 
The goal of the adversary is to obtain a local copy of a machine learning model with a higher accuracy even when the adversary owns less data than the public server of the original model~\cite{OH18,JS+19}. In general, benefits for the adversary are significant because gathering data and its training are tasks with heavy costs. Moreover, according to Juuti et al.~\cite{JS+19}, transferable adversarial examples~\cite{SL14} to analyze misidentifiable predictions via a substitute model have been discussed as applications of model extraction attacks. Consequently, a model extraction attack is a serious problem for deep learning and its underlying machine learning.

In spite of the significance of model extraction attacks, 
only simple architectures such as a logistic regression model~\cite{FFA16} or deep neural networks (DNNs) with simple architectures~\cite{JS+19,MBV18} have been discussed in existing works. Thus, the features and feasibility of model extraction attacks on other architectures are unclear. For instance, threats of model extraction attacks are non-trivial for \textit{recurrent neural networks (RNNs)} which are used in natural language processing and cybersecurity applications. In particular, the computational process of deep learning differs according to the architecture, and thus the success conditions and advantages of an adversary may differ according to the architecture as well. 
Hence, discussion on model extraction attacks for various architectures is an important research theme to avoid many potential threats as applications of the attacks, 
e.g., adversarial examples via a substitute model as described above.

\subsection{Contribution}

In this paper, we introduce model extraction attacks against RNNs and long-short term memory (LSTM) and show that RNNs with a higher prediction accuracy as a substitute model, i.e., by an adversary, can be obtained from LSTM as an original model, i.e., by a public server. Our technical contributions include finding attacks based on features of RNNs and LSTMs for a classification task and a regression task. 
Intuitively, substitute models with higher accuracies can be obtained based on the features of RNNs and LSTMs, which compute an output for each time and give feedbacks to the next time as well as considering architectures of the models and those loss functions to obtain final outputs (See Section~\ref{Experiment} for the detais). 
Despite RNNs having simpler architectures than LSTMs, an adversary can train a substitute model with a high accuracy without final outputs from an original model by biasing the outputs halfway through the sequence. 

We also conduct experiments with the MNIST dataset\footnote{\url{http://yann.lecun.com/exdb/mnist/}} for a classification task and with the Air Quality dataset\footnote{\url{https://archive.ics.uci.edu/ml/datasets/air+quality}} for a regression task to show the ability to extract RNNs as substitute models from LSTMs as original models. 
Results show that, for the MNIST dataset, the substitute model with 97.5\% prediction accuracy can be extracted only with 20\% of training data in comparison to 97.3\% accuracy of the original model. For the Air Quality dataset, the substitute model with 87.2\% accuracy, which is computed by the coefficient $R^2$ of determination, can be extracted only with training data for three months in comparison with the original model which achieves 89.9\% accuracy with training data for ten months. 
We also discuss relationships between architectures and those features and shed light on the success factors for model extraction and countermeasures (See Section~\ref{Consideration} for details).

\subsection{Related Work}

\textbf{State-of-the-Art Attacks:} 
In one of the latest results, Reith et al.~\cite{RTO19} discussed model extraction against support vector regression. 
For works targeting neural networks, Juuti et al.~\cite{JS+19} showed an attack in which an adversary generates queries for DNNs with simple architectures. 
Their attack may improve our results on RNNs, but we leave this as an open problem for RNNs. Concurrently, Wang et al.~\cite{BN18} proposed model extraction attacks to steal hyperparameters against a simple architecture, e.g., a neural network with three layers. 
To the best of our knowledge, the most elegant attack was shown by Okada and Hasegawa~\cite{OH18}. 
They utilized \textit{distillation}~\cite{CRA06,GOJ15}, which is a technique for model compression described below, to execute model extraction attacks against DNNs and convolution neural networks (CNNs) for image classification. 
In doing so, Okada and Hasegawa succeeded in extracting a model with a higher accuracy than the original model. Therefore, we consider their work as having the best results from the standpoint of the use of distillation. 

\textbf{Relationships between Architectures and Accuracy:} 
One of the main discussions in this paper is to clarify relationships between accuracy and architectures on an original model and a substitute model. 
Similar discussions have been done by Juuti et al.~\cite{JS+19}, Pal et al.~\cite{PG+19}, Krishna et al.~\cite{KT+19}, and Okada et al.~\cite{OH18}. 
The papers in ~\cite{JS+19, PG+19} explain that the accuracy of the substitute model increases in general when the architecture of an original model is identical to that of a substitute model. 
Pal et al. ~\cite{PG+19} have claimed that their argument is true unless underfitting or overfitting is caused on a substitute model, whose architecture is more complicated than that of the original model, e.g., the use of a deeper network in the substitute model than in the original model. 
In contrast, the results in~\cite{KT+19} showed that, when BERT~\cite{BERT} utilized in natural language processing is targeted, the accuracy of the substitute model is improved by the use of a deeper BERT model as the substitute model than the original model. 
The results in~\cite{OH18} also showed a similar result in a case of DNNs and CNNs. 
Our results are intuitively identical to the results in~\cite{KT+19, OH18}, except for the use of RNNs.

\textbf{Distillation:} 
In additional related works, model compression, named distillation~\cite{CRA06}, is represented. 
Distillation is used for reducing learning information by multiple neural networks, which are named as teacher models, to smaller neural networks, which are named student models. 
Hinton et al.~\cite{GOJ15} showed a method to distillate a model by a \textit{softmax function with temperature}, which can control the convergence of training through temperature. 
While distillation allows a student model to extract a large amount of information from a teacher model, model extraction attacks require that an adversary have as little as possible access to a dataset and an original. 

\textbf{Additional Features on Model Extraction:} 
As one of the latest features on model extraction attacks, the Knockoff attack~\cite{OSF19} discusses how an adversary tries a model extraction attack based solely on observed input-output pairs, i.e., without any knowledge of the dataset for an original model. However, Atli et al.~\cite{AS+19} showed that the performance of the Knockoff attack is limited. 
Meanwhile, Jagielski et al.~\cite{JC+19} proposed a new feature named fidelity to measure the general agreement between an original model and a substitute model. 
Discussions on these features on RNNs remain an open problem. 

\textbf{Further Attacks to Encourage Model Extraction:}
Naghibijouybari et al.~\cite{HAZ+18} and Yoshida et al.~\cite{YKS+19} introduced side-channel attacks targeting models separated by hardware mechanisms. Model reverse-engineering attacks~\cite{BBJ+18} where an original model is operated in an environment owned by an adversary have also been shown as advanced attacks. 
These attacks are stronger than the attack scenario in this work because we do not discuss such physical access to an original model for an adversary. Our attack against RNNs can potentially become stronger by utilizing the side-channel attack described above.

\subsection{Paper Organization}

The rest of this paper is organized as follows. 
First, the background required for understanding this paper is presented in Section~\ref{Preliminary}. 
Next, an attack model and the proposed attacks against RNNs are presented in Section~\ref{Attacks against RNN}. 
Then, experiments are shown in Section~\ref{Experiment}, and considerations including potential countermeasures are shown in Section~\ref{Consideration}. 
Finally, the conclusion and future directions are presented in Section~\ref{Conclusion}.

%% file: sec2.tex
\section{Preliminaries} \label{Preliminary}

In this section, we provide the background to understanding our work. 

\subsection{Tasks Specified in Neural Networks}

A mechanism of a neural network varies depending on the task to be solved and especially loss functions is completely different for the task. 
Hence, it is necessary to discuss attack techniques separately. 
We describe the two types of tasks handled by neural networks below.

\subsubsection{Classification} \label{sec:classification}

A classification task is a task where an input is classified into one of the categories specified in advance as an output of prediction. 
For example, a neural network for classification categorizes a number from 0 to 9 when a handwritten digit is given as input. The number of neurons in the neural network required to solve such a classification task is identical with the number of candidates in an output layer, and decides which neuron has the largest calculation result. 
The training is often executed through a softmax function to return the candidates with probabilities.

\subsubsection{Regression} \label{sec:regression} 

A regression task outputs continuous values as a response to input. For example, a neural network for the prediction of air quality outputs a numerical value of a component of the air quality at a certain time of the given input on the previous several hours. 
The neural network that solves a regression task outputs a numerical value on an output layer. Here, the neural network has plural neurons in proportion to the number of candidates predicted on the output layer. Unlike the classification task, values of neurons in the output layer are computed without a softmax function.

\subsection{Recurrent Neural Networks} \label{RNN}

\subsubsection{Principle of Recurrent Neural Networks} 

Recurrent neural networks (RNNs) are a kind of neural network that deals with time-series data including contexts, e.g., speech recognition and language processing. 
%
RNNs have negative feedbacks in those networks. 
In comparison with the architectures of typical neural networks, such as deep neural networks (DNNs), RNNs have a similar input layer, hidden layer, and output layer, but they also have a feedback path to return the output of the hidden layer to the input itself.

We describe RNNs in detail below. 
An RNN takes an input at each time. An input given at time $t$ propagates from the input layer to the hidden layer in a similar manner as in conventional neural networks. An output of the hidden layer with an activation function propagates to the output layer and, in parallel, returns to the input of the hidden layer itself as feedback. 
The signal propagated to the output layer is output as a prediction result at time $t$, whereas the feedback is given to the hidden layer as a part of an input at the next time $t + 1$. 
Consequently, the output at $t + 1$ is affected by the outputs of the hidden layer before time $t$, and hence, is able to capture contexts of time-series data. 
Unlike typical neural networks which approximate a mapping from a single input to a single output, an RNN approximates a mapping from a sequence to a sequence and returns outputs at every time. 
RNNs are used in both classification and regression tasks. In particular, RNNs often deal with a regression task to predict continuous values, such as stock price prediction~\cite{GBM+19}.

\subsubsection{Long Short-Term Memory} \label{LSTM}

In general, neural networks with deeper layers have a gradient loss problem~\cite{goodfellow2016deep}. 
Since RNNs have a recursive structure in a hidden layer, information propagates deeper when input time-series data as input becomes longer, even for shallow networks. 
Consequently, RNNs are prone to the gradient loss problem, i.e., memory is difficult to keep for a long period. 
\textit{Long short-term memory (LSTM)} has been proposed to overcome the gradient loss problem of RNNs, which only store memory for a short period. 
The basic architecture of LSTMs is the same as that of RNNs, except that a hidden layer with a recursive structure of RNNs is replaced with a layer with an element called \textit{memory unit}. 

Compared to standard RNNs, LSTMs have larger computational complexity because they have more complicated architectures. 
Furthermore, LSTMs have different types depending on the types of inputs and outputs. 
For instance, LSTMs have the following three types. The many-to-many type has inputs and outputs at each time. The many-to-one type has inputs given at each time and a single output at only the last time. The one-to-many type has a single input at only the first time and an individual output at each time. 
Typical neural networks have a single output for each given input, and hence, can be considered as the one-to-one type. 

\subsection{Model Extraction Attacks}

We briefly describe an overview of model extraction attacks below. 
Suppose that a model stored in a public server is trained using training data $D_1\subset D$. Hereafter, the trained model is called the \textit{original model}. 
General users pay the server to utilize its hosting service and give data as input via APIs of the server. This data is input to the original model, and then the original model returns prediction results to a user as a response from the service.

Based on this background, an adversary, who knows a part of the training dataset $D_2 \subset D$, executes the original model through the API to train his/her own model by utilizing prediction results from the original model for $D_2$. 
The model trained by the adversary is called \textit{substitute model}. 
For example, the adversary can train a substitute model by using prediction results and computational resources of the original model as a springboard to obtain the same or higher accuracy as that of the original model. 
This is the intuition of an attack strategy for model extraction attacks. 

The main advantage of model extraction attacks for an adversary is to obtain a model with significantly reduced costs for both data collection and its resulting training.  In general, data collection and training tasks need heavy costs, and hence resulting models become an important asset for a provider of a public server hosting an original model.  
In contrast, the adversary can obtain a substitute model whose performance can be the same as the original model without paying such expensive costs. 
Moreover, the state-of-the-art work in ~\cite{OH18} has shown that model extraction attacks enable an adversary to obtain a substitute model with higher accuracy than an original model. 

%% file: sec3.tex
\section{Model Extraction Attacks against Recurrent Neural Networks} \label{Attacks against RNN}

\subsection{Problem Setting} \label{Problem Setting}

We describe the technical problems and conditions for model extraction attacks against RNNs as the main problem setting below. 

\textbf{Computational Resources:} An original model is provided on a public server with a rich computational resource, whereas an adversary, who executes model extraction attacks, needs to train a substitute model with less resources. 

\textbf{Input/Output:} In contrast to deep neural networks (DNNs), RNNs contain an input and an output for each time. These inputs and outputs are utilized halfway through the sequences in feedback to compute for a final output. We discuss how an adversary obtains an advantage for model extraction attacks from the inputs and outputs halfway through the sequences. 

\textbf{Regression:} As described in Section~\ref{RNN}, RNNs are often utilized for a regression task. Existing model extraction attacks have been discussed mainly about classification tasks, such as image classification. A softmax function is utilized in neural networks for a classification task, but it cannot be used for a regression task. Therefore, known techniques~\cite{OH18} that modify the softmax function to decrease the number of queries cannot be used. Moreover, since an adversary who owns several parts of a dataset used in an original model may know the correct result for each output in advance, the merit of the information obtained from the APIs may downgrade in comparison with the DNNs. To effectively utilize information from RNNs, a loss function for a regression task should be constructed in detail.

In this paper, we evaluate the accuracy of model extraction attacks against RNNs from the standpoints described above. 

\subsection{Attack Strategy}

In this section, we describe model extraction attacks based on features of RNNs. 

As described in the previous section, two features, i.e., computational resource and input/output, should be considered for model extraction attacks against RNNs.
First, the architecture of an LTSM is more complicated than that of an RNN. 
Hereafter, we simply denote RNNs with a simple architecture as RNNs. 
Second, in comparison with other neural networks, such as DNNs or convolutional neural networks (CNNs), an output with variable length is generated for each time in RNNs.

Hence, we discuss model extraction attacks against RNNs from the following standpoints: 
\begin{enumerate}
  \item Can a substitute model consisting of RNN be extracted from an original model consisting of LSTM?
  \item Can an adversary obtain any advantage by utilizing features of input/output for RNNs?
\end{enumerate}
We describe the details of the attacks below. 
Let training data used in an original model be $D_c$ and training data used in a substitute model be $D_a$. 
Here, the original model, i.e., an LSTM, is trained by utilizing $D_c$. 
Then, an adversary trains the substitute model, i.e., an RNN, with $D_a$, and then continues to train RNN using the obtained prediction results from the original model by giving input data.

\subsubsection{Attack on Classification Task} 
\begin{figure}[t!]
\includegraphics[width=.50\textwidth]{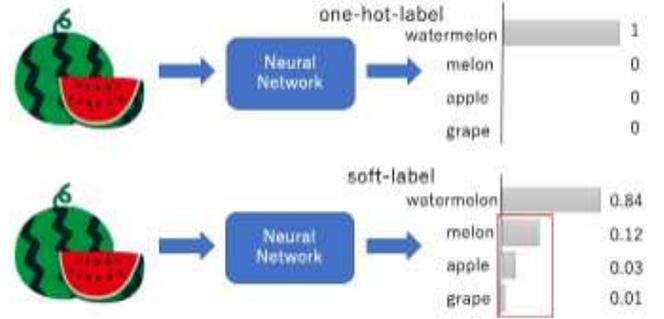}
\caption{One-Hot Encoding and Soft-Label Encoding}
\label{fig:label}
\end{figure}

\begin{figure}[t!]
\includegraphics[width=.50\textwidth]{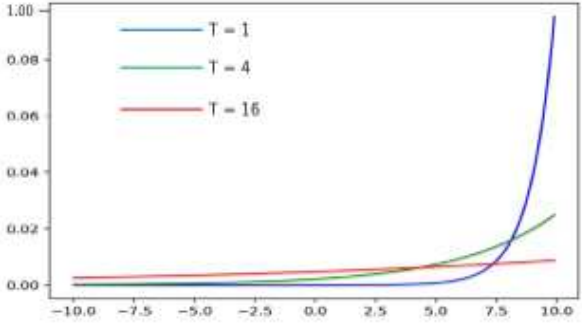}
\caption{Softmax Function with Temperature}
\label{fig:tsoft}
\end{figure}

In general, a neural network that solves a classification task returns an output without the use of a softmax function in its output layer for prediction. 
We call such an output, i.e., values not obtained through a softmax function, \textit{logits}. 
While logits are soft-label encoding where each label is output with a probability as a prediction result, \textit{labeled data} utilized in training an original model and a substitute model, i.e., $D_c$ and $D_a$, are one-hot encoding that represents just a true value for each label. 
The intuition of soft-label encoding and one-hot encoding is shown in Fig.~\ref{fig:label}. 
In this situation, an adversary executes the following attack procedure: 
\label{sec:attack_class_problem}
\begin{enumerate}
 \item \textbf{Identification of Leaky Time for the Original Model}: 
The adversary identifies the maximized index on each vector for logits returned from the original model from the first time to the last time and then evaluates the prediction accuracy for each time by comparing those labels. 
A time with high accuracy is defined as a \textit{leaky time}. 
  
  \item \textbf{Intensive Extraction at Leaky Time}: 
For the leaky time described in the previous item, the \textit{softmax function with temperature}~\cite{GOJ15} is utilized. In particular, the adversary first trains a substitute model by computing a loss function with labeled data, i.e., one-hot labels, included in $D_a$ to update the parameters. 
Then, the adversary computes the loss function, in which soft-labels through the softmax function with temperature are set as labeled data, to update the parameters. 
\end{enumerate}
The softmax function with temperature is defined in equation~(\ref{eq:tsoftmax}) and its output is shown in Fig.~\ref{fig:tsoft}. 
Intuitively, the behavior of this function is identical to the behavior of the original softmax function when $T=1$, and a gradient becomes smaller in proportion to a temperature $T$, i.e., convergence of training can become faster in proportion to $T$. 
In this paper, the softmax function with temperature is utilized and, in doing so, the effects on model extraction attacks are evaluated by changing $T$. 
\begin{eqnarray}
\label{eq:tsoftmax}
\textit{softmax}(k) = \frac{\mathrm{e}^{\frac{a_k}{T}}}{\sum_{i=1}^n \mathrm{e}^{\frac{a_i}{T}}} . 
\end{eqnarray}

\subsubsection{Attack on Regression Task}
\label{sec:lreg}

A neural network that solves a regression task returns a predicted value computed by a model, and the output is drastically different from a model for a classification task whose output is labeled with probabilities. 
Consequently, due to the structure of an output layer, a softmax function cannot be used as a loss function. 
In general, norms are utilized in a loss function for a model to solve a regression task, e.g., $L_1loss$ or $L_2loss$. 

The softmax function with temperature and $L_2loss$ have been utilized in the distillation of neural networks~\cite{LR14, GWX+17}. 
However, while the softmax function with temperature has been used in model extraction attacks by Okada and Hasegawa~\cite{OH18}, the use of $L_2loss$ in model extraction attacks is non-trivial. 
In this paper, we utilize $L_2loss$ in the model extraction attacks for a regression task.

\begin{figure}[t!]
\includegraphics[width=.50\textwidth]{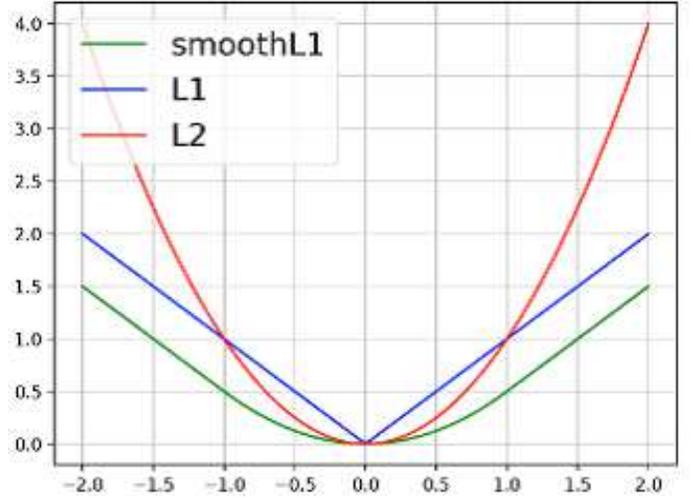}
\caption{Graphs for Loss Functions}
\label{fig:losses}
\end{figure}

While an output of the softmax function with temperature is distributed within [0,1] even for a corrupted prediction, a loss function with the norm does not have any restriction in the range of output. That is, a student model in distillation, i.e., a substitute model for a model extraction attack, may be affected greatly by a corrupted prediction from a teacher model, i.e., an original model. 
To overcome the limitation described above, instead of the use of outputs from the teacher model, we focus on the method by Chen et al.~\cite{GWX+17} which utilizes the outputs as an upper bound to be achieved for the student model. 
As shown in equation~(\ref{eq:lb}), a penalty to adjust a parameter is given for the output only when $L_2loss$ between a predicted value $R_t$ for the teacher model and labeled data $y$ is smaller than $L_2loss$ between a predicted value $R_s$ for the student model and $y$ for the teacher model with respect to a parameter $m$ designated in advance. 
The function is called \textit{teacher bound regression loss} and is denoted by $L_b$. 
Here, $m$ is a small value estimated by distribution for each dataset. 
\begin{eqnarray}
\label{eq:lb}
L_b = \left\{
    \begin{array}{l}
      || R_s - y ||_2^2, if ||R_s - y ||_2^2 + m > || R_t -y ||_2^2 \\
      0, otherwise.
    \end{array}
  \right.
\end{eqnarray}
Furthermore, instead of the use of $L_b$ as a loss function for the student model, $L_b$ is embedded in the use of \textit{smooth L1 loss}, $L_{s1}$, which is defined in equation~(\ref{eq:ls1}). 
The smooth L1 loss $L_{s1}$ can overcome the problem where the derivation is impossible by closing to zero on the L1 loss and the problem where a gradient becomes too large in proportion to the distance on L2. 
The distribution for each loss function is shown in Fig.~\ref{fig:losses}. 
\begin{eqnarray}
\label{eq:ls1}
L_{s1} = \left\{
    \begin{array}{l}
      0.5| R_s - y |_1^2, if | R_s - y |_1 < 1\\
      | R_s - y |_1 - 0.5, otherwise.
    \end{array}
  \right.
\end{eqnarray}
Then, the loss function $L_{reg}$ of a student model for a regression task is defined as follows: 
\begin{eqnarray}
\label{eq:lreg}
L_{reg} = L_{s1}(R_s, y) + vL_b .
\end{eqnarray}
For model extraction attacks, the teacher model described above is dealt with by an original model stored in a public server while the student model is dealt with a substitute model. 
In doing so, $L_{reg}$ is utilized as a loss function to train the substitute model. 
Since an adversary does not know the outputs identical to the input data on typical neural networks, an output returned from an original model is important information for the adversary. Hence, discussion about the details in loss functions described above is often unnecessary. 

Meanwhile, in the case of RNNs, the following discussion is necessary due to features of their input and output. 
For RNNs with the many-to-many type, an adversary can correctly guess the predicted data returned from an original model through its own data because both input data and output data are included in the same dataset as time-series data.  
In other words, an adversary who owns a part of a dataset can train a substitute model with its own input and labeled data. 
In doing so, to give a more generalized performance to the substitute model, knowledge of the original model trained with a larger amount of data should be extracted by the adversary. 
We evaluate the loss function $L_{s1} $ with respect to the extraction of knowledge described above.

%% file: sec4.tex
\section{Experiment} \label{Experiment}

In this section, we conduct experiments on model extraction attacks against the RNN described in the previous section to evaluate their effectiveness in terms of the accuracy of prediction in measuring the correctness of predictions on the test distribution. 
In particular, we discuss neural network architectures for both classification and regression tasks. 

\subsection{Experiment Setup} 
The experimental environment is shown in Table~\ref{tab:exp_env}. 
We configured the environment on the Google Colaboratory\footnote{\url{https://colab.research.google.com}}. 
The training algorithm of neural networks used in the experiments is the Adam optimizer, which is standard equipment for TensorFlow\footnote{\url{https://www.tensorflow.org/}} with a learning rate of 0.001.  
\begin{table}[t!]
\centering 
\caption{Experimental Environment}
\label{tab:exp_env}
  \begin{tabular}{|c|c|} \hline
    Development Platform & Tensor Flow 2.0. \\ \hline
    OS & Ubuntu 18.04 \\ \hline 
    GPU & NVIDIA Tesla K80 12GB \\ \hline
    Memory & 13GB RAM  \\ \hline
    Storage & 360GB \\ \hline
  \end{tabular}
\end{table}

\subsubsection{Setting for the Classification Task}
\begin{figure}[t!]
\centering
\includegraphics[height=0.1\textheight, width=0.1\textwidth]{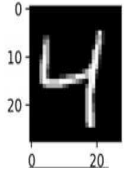}
\caption{Example of MNIST Dataset}
\label{fig:mnist_datasets}
\end{figure}

We utilize the MNIST dataset\footnote{\url{http://yann.lecun.com/exdb/mnist/}} in an experiment on model extraction attacks against a many-to-many LSTM. 
The MNIST dataset used consists of 55,000 samples as training data and 11,000 samples as test data. 
Each sample represents a handwritten character from 0 to 9 and 
is represented as $28 \times 28$ pixels. Fig.~\ref{fig:mnist_datasets}\footnote{\url{https://machinelearningmastery.com/how-to-develop-a-convolutional-} \url{neural-network-from-scratch-for-mnist-handwritten-digit-classification/}} shows examples of the samples on the MNIST dataset.

\begin{figure}[t!]
\includegraphics[width=.50\textwidth]{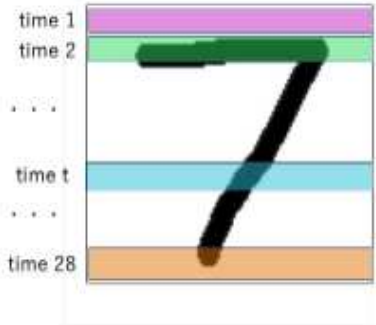}
\caption{Conversion of MNIST Dataset to Time-Series Data}
\label{fig:mnist_time}
\end{figure}

In the experiment described below, each sample on the MNIST dataset is converted into time-series data. As shown in Fig.~\ref{fig:mnist_time}, each sample of $28 \times 28$ pixels is divided into lines for each time sequentially, where we suppose the $t$-th line is input at time $t$. That is, there are 28 lines of input data, and each line is given to a model as time-series data for times 1 to 28. 

As neural networks for classification of handwritten digits, 
DNNs and CNNs have been discussed by Okada and Hasegawa with respect to model extraction attacks based on the softmax function with temperature. 
To compare with the work by Okada and Hasegawa, we adopt the same experimental setting as in their work. 
In particular, 
55,000 samples as training data in the MNIST dataset are divided into five subsets, i.e., 11,000 samples per subset. 
Let the subsets be denoted as $ D_1 $, $ D_2 $, $ D_3 $, $ D_4 $, and $ D_5 $ for convenience. 
Four of the subsets are utilized as training data $ D_c $ for an original model, and the remaining subset is utilized as training data $ D_a $ for a substitute model by an adversary. 
Experiments are conducted five times because there are five cases in which each subset is used as $ D_a $, and then the final results of the entire experiment are the averages of the results of the five experiments, i.e., 5-fold cross validation. 

Since the MNIST dataset is an academic benchmark for a classification task, as described in Section~\ref{sec:attack_class_problem}, we utilize cross-entropy error through the softmax function with temperature as a loss function. Here, let temperatures be $ T = 1, 4, 16 $. For $ T = 1 $, the softmax function with temperature is exactly identical to the original softmax function as described above. 
After training the substitute model with labeled data using the cross-entropy error, the substitute model is trained as a loss function utilizing the softmax function with temperature to logits obtained from the original model. 

The accuracy of prediction is evaluated with 10,000 samples as test data of the MNIST dataset with respect to both the original model of LSTM and the substitute model of RNN. 
According to our pre-experiment, the accuracy of the original model is 97.3\% although we omit the details. The accuracy is the attack goal of the substitute model for an adversary. 
The setting for training in this experiment is shown in Table~\ref{tab:exp_learning_class}. 
\begin{table}[t!]
\centering 
\caption{Setting for Training in Case of Classification Task}
\label{tab:exp_learning_class}
  \begin{tabular}{|c|c|} \hline
    Epoch & 220 \\ \hline 
    Iterations in Each Epoch & $\frac{|D_a|}{50}$ \\ \hline
    Batch Size &  50 \\ \hline
  \end{tabular}
\end{table}

\begin{figure}[t!]
\includegraphics[width=.50\textwidth, height=.25\textheight]{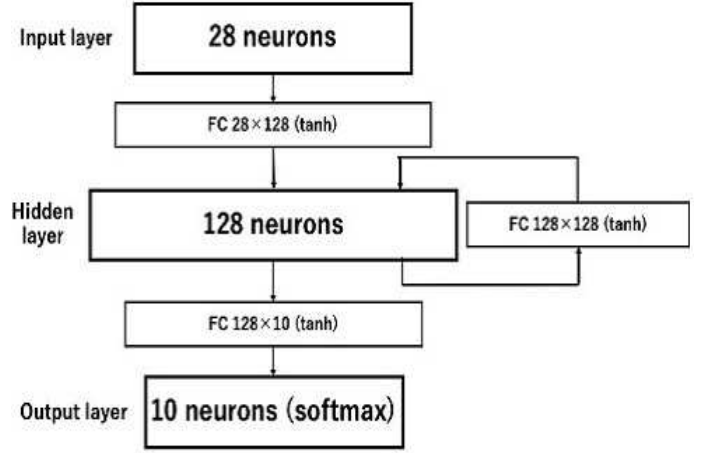}
\caption{Architecture of Neural Network Used on the Experiment for Classification Task}
\label{fig:class_layers}
\end{figure}

Fig.~\ref{fig:class_layers} shows an architecture of the neural network used on the experiment for classification task.
The hidden layer is RNNcell. The outputs of the RNNcell neurons at time $t$ and the inputs to the RNNcell neurons at time $t + 1$ are connected by a full connection, i.e., FC. We use \textit{tf.contrib.rnn.BasicLSTMCell} and \textit{tf.contrib.rnn.BasicRNNCell} as a hidden layer of RNN and LSTM, respectively, and use \textit{tf.contrib.rnn.static\_rnn} for network input/output.

\subsubsection{Settings for Regression Task}

\begin{figure}[t!]
\includegraphics[width=.5\textwidth, height=.25\textheight]{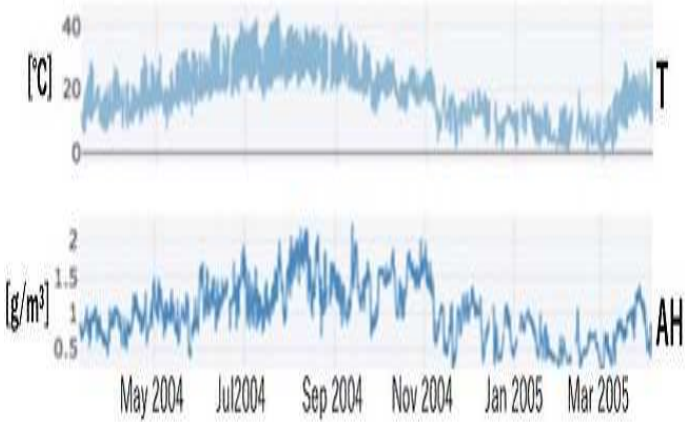}
\caption{Examples of the Air Quality dataset}
\label{fig:aqdt}
\end{figure}

For experiments on model extraction attacks against the LSTM to solve the regression task, we deal with an Air Quality dataset\footnote{\url{https://archive.ics.uci.edu/ml/datasets/air+quality}}, which consists of the amount of materials contained in the atmosphere and temperatures collected by sensor devices. 
The Air Quality dataset was measured every hour from 18:00 on 03/10/2004 to 14:00 04/04/2005 and each record per hour consists of 13 kinds of values as the amount of materials and temperature. Examples of the Air Quality dataset are shown in Fig.~\ref{fig:aqdt}\footnote{\url{https://www.atmarkit.co.jp/ait/articles/1804/26/news150.html}}   

A dataset that minimizes missing values is desirable because we handle time-series data, ajd thus a reliable experimental result can be expected by using the Air Quality dataset as follows. We utilize six values, i.e., temperature $T$, absolute humidity $AH$, time average value $CO$ of oxide for carbon monoxide, time average value $NMHC$ of titanium oxide for non-methane hydrocarbons, time average value $NO_x$ of tungsten oxide for nitrogen oxide, and time average value $NO_2$ of tungsten oxide for nitrogen dioxide. 
In this experiment, we let an RNN receive the six values described above for 72 hours as time-series data of input and then we predicted those values as values measured on the 73rd hour as output. 

We separated the Air Quality dataset for test data, i.e., data in 2005, and training data, i.e., data in 2004 data, as a typical setting. Let $D_c$ be the training data for an original model, where an adversary has training data $D_a \in D_c$ on some $n$ months. 
Because there are multiple ways to select $D_a$, experiments are conducted according to Table~\ref{tab:exp_env2} and then the final result of the entire experiment is the average of results on all the experiments. 
\begin{table}[t!]
\centering 
\caption{Combination of Training Data Owned by Adversary}
\label{tab:exp_env2}
  \begin{tabular}{|l|c|r||r|} \hline
    Scale of Training Data & Month of Training Data Used in Experiment \\ \hline
    3 Months & \{4, 5, 6\}, \{7, 8, 9\}, \{10, 11, 12\} \\ \hline
    6 Months & \{4, 5, 6, 7, 8, 9\}, \{7, 8, 9, 10, 11, 12\} \\ \hline
  \end{tabular}
\end{table}

As described above, the softmax function with temperature cannot be used in the regression task. Therefore, the $L_{reg}$ described in Section~\ref{sec:lreg} is used instead. 
Note again that $L_{reg}$ is used in the distillation of neural networks as well as the softmax function with temperature, and thus it is expected to be effective in model extraction attacks. 
A substitute model is trained 10,000 times using $L_2loss$ of the predicted values and labeled data as $D_a$, and then it is trained 10,000 times again by  using $L_{reg}$ with the predicted values via querying to the original model. 
The batch size is 16.

The coefficient $R^2$ of determination, which is commonly used for regression analysis, is utilized in the evaluation of the accuracy. 
$R^2$ is defined as follows: 
\begin{eqnarray}
\label{eq:r2}
R^2 = 1- \frac{\sum_{i=1}^n (y_i - pre)^2}{\sum_{i=1}^n (y_i - \overline{y})^2}, 
\end{eqnarray}
where $y$ is the labeled data, $\overline{y}$ is the average of $y$, and $pre$ are the predictions from the original or substitute model. 
Intuitively, $R^2$ is desirably very close to 1. 
According to Rubin~\cite{Rubin12}, 
if $R^2$ exceeds 0.8, the correlation is very strong and, for example, $R^2 = 0.8$ means that 80\% of the variation in the dependent variable has been explained. 
Based on our pre-experiment, $R^2$ of the original model is 0.8992 although we omit the details. In this experiment, we refer to the value of $R^2$ as the accuracy, and an $R^2$ of 0.8992 is the attack goal of the substitute model for an adversary.

An architecture of the neural network used on this experiment is almost the same as Fig.~\ref{fig:class_layers} except that a loss function in the output layer and the number of neurons for each layer and are different. 
In particular, the loss function consists of the $L_{reg}$ function as described above. 
Meanwhile, the input layer and output layer contain six neurons and the hidden layers contain twenty neurons, where the neurons in each layers are fully connected in the same manner as Fig.~\ref{fig:class_layers}. 
Meanwhile, 
We use \textit{tf.nn.rnn\_cell.BasicRNNCell} and \textit{tf.nn.rnn\_cell.BasicLSTMCell} as hidden layers and use \textit{tf.nn.dynamic\_rnn} for network input/output. 

\subsubsection{Evaluation Terms in Experiments} 

An evaluation of the experiments is conducted as described below. 

\paragraph{Classification Task}
The evaluation terms in the experiments for the classification task are as follows:
\begin{itemize}
  \item Leaky time. 
  \item Difference in accuracy depending on each architecture of the substitute model.  
  \item Difference in accuracy depending on the number of training data owned by an adversary, i.e., $D_a$. 
  \item Difference in accuracy depending on temperature $T$ of the softmax function with temperature. 
\end{itemize}

\paragraph{Regression Task}
The evaluation terms in the experiments for the regression task are as follows:
\begin{itemize}
  \item Difference in $R^2$ depending on each architecture of the substitute model.
  \item Difference in $R^2$ depending on the ratio of training data owned by an adversary, i.e., $D_a$. 
  \item Effects of loss functions using $L_b$.
\end{itemize}

\subsection{Experimental Result}
\subsubsection{Results of Classification Task}
\begin{figure}[t!]
\includegraphics[width=.45\textwidth, , height=.25\textheight]{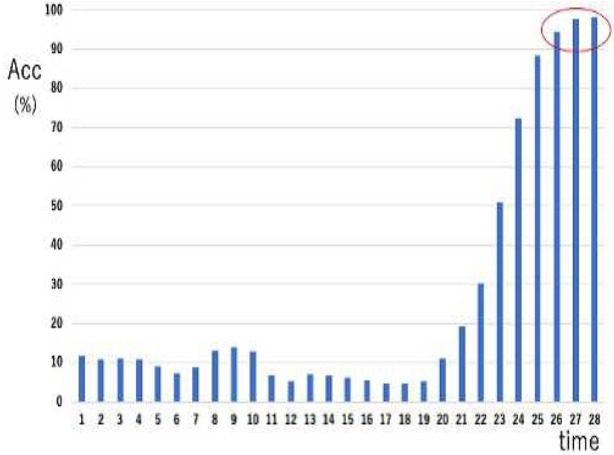}
\caption{Results for Identification of Leaky Time on LSTM}
\label{fig:leak-lstm}
\end{figure}

\begin{figure}[t!]
\includegraphics[width=.45\textwidth, height=.25\textheight]{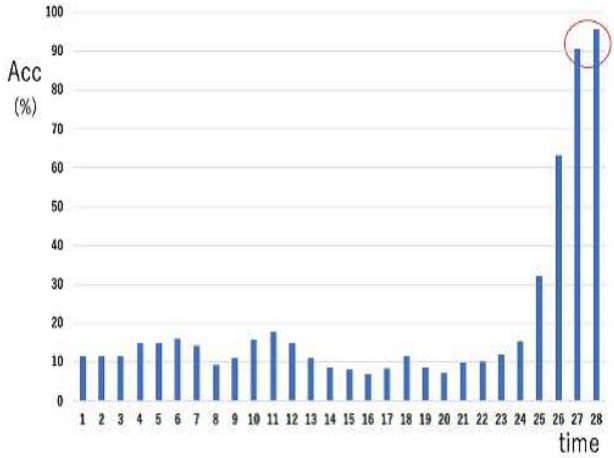}
\caption{Results for Identification of Leaky Time on RNN}
\label{fig:leak-rnn}
\end{figure}

First, the results for identification of leaky time are shown in Figs.~\ref{fig:leak-lstm} and ~\ref{fig:leak-rnn}. 
\begin{figure}[t!]
\includegraphics[width=.45\textwidth, height=.25\textheight]{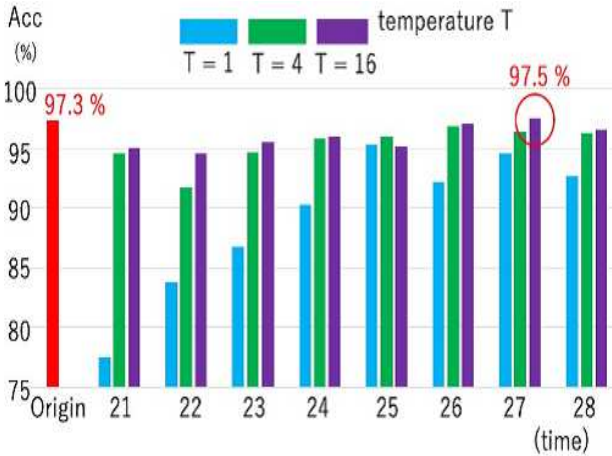}
\caption{Intensive Extraction on Leaky Times}
\label{fig:soft_extraction}
\end{figure}

According to Fig.~\ref{fig:leak-lstm} and Fig.~\ref{fig:leak-rnn}, increments of accuracy for LSTM started from $t=21$ and become more than 90\% after $t=26$. On the other hand, in an additional experiment, whereby RNN is utilized in the original model instead of LSTM, increments of accuracy for RNN started from $t=25$ and become more than 90\% after $t=27$. 
Therefore, the accuracy of the final results of LSTM is greater than that of RNN, and LSTM is potentially leakier even at an early time. 

Next, the results for intensive extraction of LSTM at $t=21$ and later times are shown in Fig.~\ref{fig:soft_extraction}. 
According to Fig.~\ref{fig:soft_extraction}, a model with 95\% accuracy is extracted at $t=21$. Notably, for $T=16$ in the softmax function with temperature, the accuracy becomes 97.5\%, which is higher than the 97.3\% accuracy of the original model. 
Note that, in the case in which the original model is a many-to-one type of LSTM and only the final prediction result at $t=27$ is returned, 
an adversary will only know the final result and cannot extract 97.5\% accuracy for the substitute model.

\begin{figure}[t!]
\includegraphics[width=.40\textwidth]{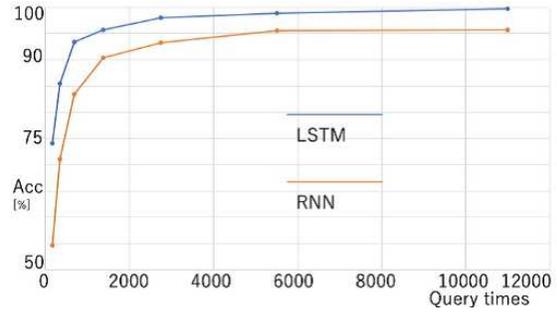}
\caption{Accuracy of Substitute Model Depending on the Number of Queries}
\label{fig:extraction_query}
\end{figure}

Finally, we investigate how the accuracy of the substitute model is related to the number of queries at time $t=27$ and temperature $T=16$. 
The result is shown in Fig.~\ref{fig:extraction_query}. 
As a result, accuracy of the substitute model is rapidly downgraded when the amount of training data $D_a$ owned by an adversary decreases.

\subsubsection{Results of the Regression Task}

\begin{figure*}[t!]
\includegraphics[width=\textwidth, height=.25\textheight]{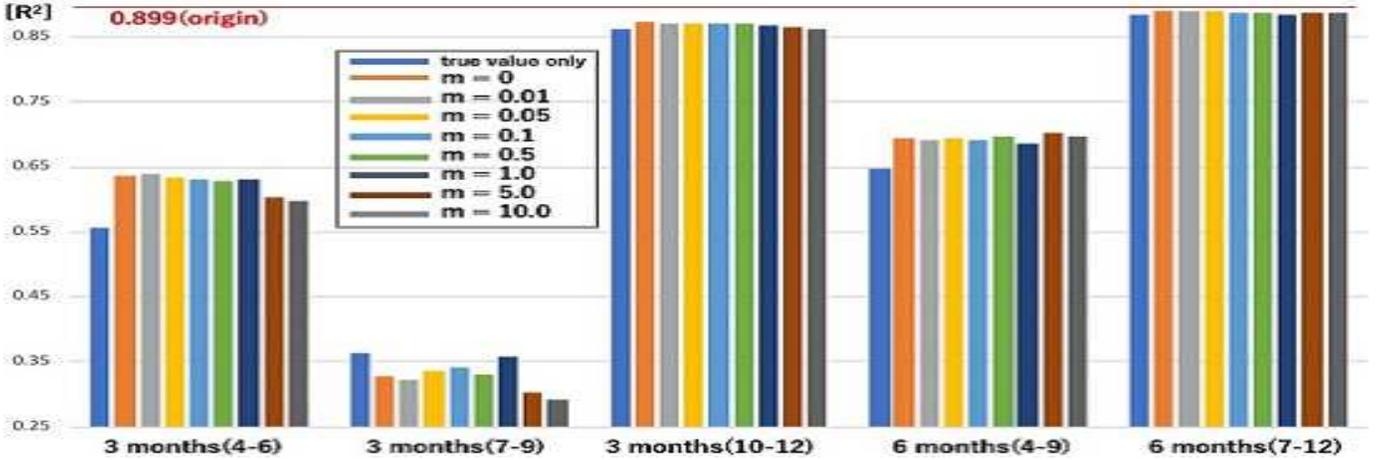}
\caption{Accuracy of Substitute Model with Air Quality Dataset}
\label{fig:regRNN}
\end{figure*}

The results with the Air Quality dataset are shown in Fig.~\ref{fig:regRNN}. When the parameter $m$ exceeds $10$, in Equation~\ref{eq:lb}, $|| R_s - y ||_2^2$ is selected as $L_b$ almost every time and $m$ does not cause a change in the experimental results, and thus we show results where $m$ is $10$ or less.
\begin{figure*}[t!]
\includegraphics[width=\textwidth, height=.25\textheight]{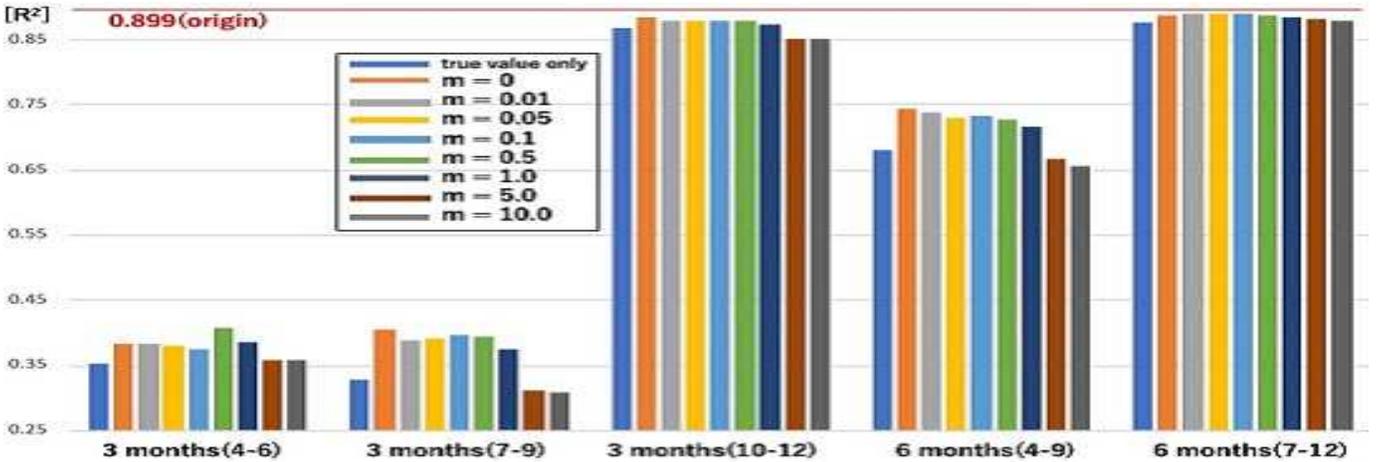}
\caption{Accuracy in Air Quality Dataset When Adversary Owns LSTM}
\label{fig:regLSTM}
\end{figure*}
According to Fig.~\ref{fig:regRNN}, in comparison with a substitute model trained with only labeled data without querying to the original model, a substitute model with the loss function $L_{reg}$, i.e., in equation (\ref{eq:lreg}), using the predicted values from the original model provides a large value for $R^2$. 
The results imply that the parameter $m$ becomes irrelevant when it is larger than $5$.
Besides, when an adversary has data for three months from July to September, even the use of the loss function is ineffective if training for the substitute model with labeled data is unsuccessful. 
Moreover, $R^2$ does not depend on $m$ when the model extraction attacks are performed with high performance, e.g., $R^2$ exceeds 0.85.

Next, we conduct an experiment where an adversary owns an LSTM as a substitute model. 
The results are shown in Fig.~\ref{fig:regLSTM}. 
According to Figs.~\ref{fig:regRNN} and ~\ref{fig:regLSTM}, the effects of the loss functions are independent of the use of LSTM and RNN. 
Meanwhile, an adversary utilizing LSTM can extract a substitute model with a higher accuracy because the use of LSTM makes $R^2$ larger. 
In comparison with the use of RNN as a substitute model, the value of $R^2$ is much lower for data from the three months of April to June because parts of the Air Quality dataset are unfit for LSTM. 
Similar to the results in the previous paragraph, the use of the loss function is ineffective if training for a substitute model is unsuccessful. 

The overall results show that a substitute model with a high accuracy can be extracted even when an adversary owns a small amount of data and provides a number of queries, e.g., three months from October to December,.

%% file: sec5.tex
\section{Considerations} 
\label{Consideration}

In this section, we discuss the differences in the behaviors of model extraction attacks according to the distinction of architecture in the classification problem to determine how simple RNNs and complex LSTMs affect model extraction attacks. Moreover, we consider the loss function used in the regression problem to clarify how the designed loss function works in the framework of model extraction attack, and then show future prospects. Finally, we discuss countermeasures against model extraction attacks.

\subsection{Classification Problem}
We discuss the impact of structural architectural differences in the original model. 
LSTM can stochastically control information not 
only on a near past but also an earlier time that can be fielded back, and thus a range of leaky times can be potentially expanded. 
Likewise, a substitute model with a high accuracy at $t = 21$ was extracted because the bottom lines of the samples on the MNIST dataset contain less information. Thus, the substitute model was able to learn sufficient knowledge from the original model at the procedure on the 21st line. 
In our experiment, we converted the MNIST dataset to time-series data to identify the range of leaker times. 
This implies the existence of other time-series datasets where an adversary can extract a model even at earlier times, e.g., at the beginning time of a model extraction attack. 
In such a case, extraction becomes easier and thus more stringent restrictions in the use of APIs are necessary.

Moreover, as an additional experiment, we used LSTM as a substitute model. 
In comparison with the 90.3\% accuracy obtained by RNN as a substitute model with 1,375 training data, the same accuracy is obtained by LSTM as a substitute model with only 500 training data. 
Furthermore, in the case of LSTM as a substitute model, the model with 99.69\% accuracy can be extracted when 11,000 training data is used in the substitute model. Thus, in RNNs, an adversary using LSTM can extract a substitute model with a higher accuracy even with fewer queries than the use of RNN. 

Okada et al.~\cite{OH18} utilized DNN and CNN as substitute models with the MNIST dataset, respectively, in their experiments. 
In doing so, they showed that models with more than 90\% accuracies can be extracted even with 684 training data on DNN and 171 training data on CNN for the substitute models. In other words, according to Okada et al., the use of CNN is more effective for an adversary to extract a model with a high accuracy in an image classification task. 
Likewise, Krishna et al.~\cite{KT+19} showed that an adversary can obtain a higher accuracy when a more complicated architecture of BERT~\cite{BERT} is used.
In contrast, we show that in the case of time-series data, the use of LSTM as a substitute model enables an adversary to extract a model with a higher accuracy even with fewer queries than the use of a RNN. Our results are identical to those of Okada et al. and Krishna et al., except for differences in the architectures among RNN, CNN, and BERT.

Finally, although a substitute model in our experiment was trained by utilizing labeled data and soft labels separately, Chen et al.~\cite{GWX+17} defined a loss function using both at the same time. The use of this loss function can reduce the number of epochs in training. In particular, the loss function is defined in quation~(\ref{eq:losskousatu}) below, 
where $ P_s $ is the predicted value of a substitute model, $ P_t $ is a predicted value of an original model, $ \mu $ is a hyperparameter, $ L_{hard} $ is the value of a loss function obtained by labeled data, and $ L_{soft} $ is the value of a loss function obtained by soft labels.
\begin{eqnarray}
\label{eq:losskousatu}
L = \mu L_{hard}(P_s, y) + (1 - \mu)L_{soft}(P_s, P_t) . 
\end{eqnarray}
The convergence of a substitute model with high accuracy becomes more effective by setting an appropriate value for $ \mu $. 
We leave the determination of an appropriate value for $\mu$ as an open problem.

\subsection{Regression Task}
We consider the loss function $L_b$. We discuss the reason why performance of a substitute model is independent of a parameter $m$ when the model extraction attacks are performed with high performance, e.g., $R^2$ exceeds 0.85. 
The parameter $m$ affects the case where a non-zero value is selected for $L_b$, i.e., frequency of training such that weights are significantly updated. 
Since training for the substitute model by an adversary was sufficiently converged in our experiment given a sufficient number of epochs, the case described above did not become a problem. 

Next, we consider the reason why the loss function $L_{reg}$ defined in equation~(\ref{eq:lreg}) is ineffective in the case where training for the substitute model with labeled data is unsuccessful. 
In that case, at the phase of training with the predict values from an original model, a value of $R_s$ is expected to be close enough to that of $R_t$. 
Consequently, if $R_s$ is mismatched and underfitting at the phase of $L_b$ in equation~(\ref{eq:lb}), the inequality cannot be evaluated as expected. 
Thus, its resulting loss function $L_{reg}$ will become unworkable.

Finally, if an adversary who owns data from April to June can produce valid data from July to September, then it has the same ability as an adversary who originally owns data from April to September. 
This is possible for at least time-series data such as the Air Quality dataset, in which an output value at a certain time is identical to an input at a later time. 
For instance, if an adversary who owns data from April to June makes queries from 0:00 on June 28th to 23:00 on June 30th, then it may be able to precisely predict data about 0:00 on July 1st, which is unknown for the adversary. 
By iterating such an operation recursively, the adversary can obtain a larger and pseudo dataset. 
We plan to verify the validity of the methodology in a future work by utilizing features of input/output for RNNs to handle time-series data.

\subsection{Countermeasures}
As countermeasures to model extraction attacks, Kesarwani et al.~\cite{MBV18} proposed extraction warning, wherein a model trained by an adversary is emulated as another model by a proxy and extraction will be alerted if the emulated model achieves some threshold designated in advance. 
Although the extraction warning can potentially be useful, settings about thresholds have never been discussed. 
In addition, although there is a method to detect model extraction attacks~\cite{JS+19} whereby a cloud sever checks if the distribution of API queries deviates from general and honest users, such an approach is ineffective against collusion between adversaries who execute the attacks according to Kesarwani et al.~\cite{MBV18}. Moreover, according to Atli et al.~\cite{AS+19}, approaches to monitoring and alerting behavior of users are ineffective against model extraction attacks on complex neural networks. 

Szyller et al.~\cite{SBS+19} proposed an approach based on digital watermarking~\cite{ZBHRV16,CLEKS19} to claim cloud's ownership after a model extraction attack is made. 
Such an approach is expected to detect model extraction attacks by verifying the watermarking in a substitute model when an adversary publishes the model. However, the model extraction attacks based on distillation shown in this work (and the work of Okada et al.~\cite{OH18}) enable an adversary to remove watermarking. Moreover, according to Krishna et al.~\cite{KT+19}, watermarking can only verify whether an original model has been stolen through a substitute model and not prevent the extraction itself. Therefore, extraction can go unnoticed if an adversary keeps its own substitute model private, making the use of digital watermarking insufficient.

Another countermeasure is the use of differential privacy~\cite{D06} as proposed by Huadi et al.~\cite{HQH+19}. 
However, such an approach will downgrade performance of an original model hosted by a cloud server. 
Consequently, to the best of our knowledge, a proposal of practical and effective countermeasures remains an open problem. 

Finally, we discuss an alternative way to mitigate model extraction attacks, especially against the attacks discussed in this paper. 
The use of the softmax function with temperature~\cite{GOJ15} was the key idea regardless of the architecture from the viewpoint of model extraction attacks for a classification task. 
In doing so, logit provides the probability for each label as a predicted result and is given to the softmax function with temperature as input. Since the accuracy heavily depends on the maximized value in the prediction result, we consider manipulating the effects by the second maximized value and the lower values to mitigate the softmax function with temperature. 
Intuitively, this can prevent an adversary from obtaining more information than prediction results alone by disturbing an output without affecting a precisely predicted result.

%% file: sec6.tex
\section{Conclusion} \label{Conclusion}

Model extraction attacks enable an adversary to extract a machine learning model via prediction queries to a model. In this paper, we discussed model extraction attacks based on features of recurrent neural networks (RNNs). 
In a case of a classification task, we extracted a substitute model without the final output from a original model by utilizing outputs halfway through the sequence. In a case of a regression task, we presented a new attack by newly configuring a loss function. 

In a classification task, we conducted experiments by converting the MNIST dataset to time-series data. Our experimental results show that a substitute model can be effectively extracted by utilizing prediction results from the original model after identifying a leaky time via training with true labels. 
In particular, by utilizing the softmax function with temperature~\cite{GOJ15} in predicted results from the original model, a substitute model with a higher prediction accuracy than the original model could be extracted. 

In a regression task, we proposed a new extension of model extraction attacks by using a teacher bounded regression loss function~\cite{GWX+17} as a loss function. 
In experiments with the Air Quality dataset for the proposed attack, we showed a substitute model whose correlation is strongly similar to an original model even when the substitute model was trained with only a small amount of data. 

We also considered relationships between the accuracy and complicated architectures for a substitute model. 
We conclude that the use of a complex architecture contributes to obtaining a higher accuracy for a substitute model. 
These results corroborate the findings of Okada and Hasegawa~\cite{OH18} and Krishna et al.~\cite{KT+19}.

We plan to extract a model with a higher accuracy in a regression task by generating new data from time-series data owned by an adversary in a future work. 
We also plan to discuss the threshold of restrictions in the use of APIs as a countermeasure to model extraction attacks.

%% file: main.bbl
\begin{thebibliography}{10}

\bibitem{FFA16}
Florian Tram{\'e}r, Fan Zhang, and Ari Juels.
\newblock Stealing machine learning models via prediction apis.
\newblock In {\em Proc. of USENIX Security}, pages 601--618. USENIX
  Association, 2016.

\bibitem{OH18}
Rina Okada and Satoshi Hasegawa.
\newblock Gazoubunruishinsogakushukinitaisuru model extraction kogekinokensho.
\newblock In {\em Proc. of CSS (in Japanese)}, pages 201--208, 2018.

\bibitem{JS+19}
Mika Juuti, Sebastian Szyller, Samuel Marchal, and N.~Asokan.
\newblock Prada: Protecting against dnn model stealing attacks.
\newblock In {\em Proc. of EuroS\&P}, pages 512--527. IEEE, 2019.

\bibitem{SL14}
Nedim {\^S}rndi{\'c} and Pavel Laskov.
\newblock Practical evasion of a learning-based classifier: A case study.
\newblock In {\em Proc. of IEEE S\&P}, pages 197--211. IEEE, 2014.

\bibitem{MBV18}
Manish Kesarwani, Bhaskar Mukhoty, Vijay Arya, and Sameep Mehta.
\newblock Model extraction warning in mlaas paradigm.
\newblock In {\em Proc. of ACSAC}, pages 371--380. ACM, 2018.

\bibitem{RTO19}
Robert Nikolai~Reith, Thomas Schneider, and Oleksandr Tkachenko.
\newblock Efficiently stealing your machine learning models.
\newblock In {\em Proc. of WPES}, pages 198--210. ACM, 2019.

\bibitem{BN18}
Binghui Wang and Neil Zhenqiang~Gong.
\newblock Stealing hyperparameters in machine learning.
\newblock In {\em Proc. of IEEE S\&P}, pages 36--52. IEEE, 2018.

\bibitem{CRA06}
Cristian Bucil{\^a}, Rich Caruana, and Alexandru Niculescu-Mizil.
\newblock Model compression.
\newblock In {\em Proc. of KDD}, pages 535--541. ACM, 2006.

\bibitem{GOJ15}
Geoffrey Hinton, Oriol Vinyals, and Jeff Dean.
\newblock Distilling the knowledge in a neural network.
\newblock 2015.

\bibitem{PG+19}
Soham Pal, Yash Gupta, Aditya Shukla, Aditya Kanade, Shirish Shevade, and Vinod
  Ganapathy.
\newblock A framework for the extraction of deep neural networks by leveraging
  public data.
\newblock 2019.

\bibitem{KT+19}
Kalpesh Krishna, Singh~Tomar Gaurav, P.~Ankur Parikh, Nicolas Papernot, and
  Mohit Iyyerm.
\newblock Thieves on sesame street! model extraction of bert-based apis.
\newblock 2019.

\bibitem{BERT}
Jacob Devlin, Ming-Wei Chang, Kenton Lee, and Kristina Toutanova.
\newblock Bert: Pre-training of deep bidirectional transformers for language
  understanding.
\newblock 2019.

\bibitem{OSF19}
Tribhuvanesh Orekondy, Bernt Schiele, and Mario Fritz.
\newblock Knockoff nets: Stealing functionality of black-box models.
\newblock In {\em Proc. of CVPR}, pages 4954--4963. IEEE, 2019.

\bibitem{AS+19}
Buse Gul~Atli, Sebastian Szyller, Mika Juuti, Samuel Marchal, and N.~Asokan.
\newblock Extraction of complex dnn models: Real threat or boogeyman?
\newblock 2019.

\bibitem{JC+19}
Matthew Jagielski, Nicholas Carlini, David Berthelot, Alex Kurakin, and Nicolas
  Papernot.
\newblock High-fidelity extraction of neural network models.
\newblock 2019.

\bibitem{HAZ+18}
Hoda Naghibijouybari, Ajaya Neupane, Zhiyun Qian, and Nael Abu-Ghazaleh.
\newblock Rendered insecure: Gpu side channel attacks are practical.
\newblock In {\em Proc. of CCS}, pages 2139--2153. ACM, 2018.

\bibitem{YKS+19}
Kota Yoshida, Takaya Kubota, Mitsuru Shiozaki, and Takeshi Fujino.
\newblock Model-extraction attack against fpga-dnn accelerator utilizing
  correlation electromagnetic analysis.
\newblock In {\em Proc. of FCCM}, pages 318--318. IEEE, 2019.

\bibitem{BBJ+18}
Lejla Batina, Shivam Bhasin, Dirmanto Jap, and Stjepan Picek.
\newblock Csi neural network: Using side-channels to recover your artificial
  neural network information.
\newblock In {\em Proc. of USENIX Security}, pages 515--532. USENIX
  Association, 2018.

\bibitem{GBM+19}
Achyut Ghosh, Soumik Bose, Giridhar Maji, C.~Narayan Debnath, and Soumya Sen.
\newblock Stock market prediction using lstms.
\newblock In {\em Proc. of CAINE 2019}, volume~63 of {\em EPiC Series in
  Computing}, pages 101--110. EasyChair, 2019.

\bibitem{goodfellow2016deep}
Ian Goodfellow, Yoshua Bengio, and Aaron Courville.
\newblock {\em Deep learning}.
\newblock MIT press, 2016.

\bibitem{LR14}
Jimmy Ba and Rich Caruana.
\newblock Do deep nets really need to be deep?
\newblock In {\em Proc. of NIPS}, pages 2654--2662. Curran Associates, Inc.,
  2014.

\bibitem{GWX+17}
Guobin Chen, Wongun Choi, Xiang Yu, Tony Han, and Manmohan Chandraker.
\newblock Learning efficient object detection models with knowledge
  distillation.
\newblock In {\em Proc. of NIPS}, pages 742--751. Curran Associates, Inc.,
  2017.

\bibitem{Rubin12}
Allen Rubin.
\newblock {\em Statistics for Evidence Based Practice and Evaluation},
  chapter~13.
\newblock Brooks/Cole Pub Co., 2012.

\bibitem{SBS+19}
Sebastian Szyller, Buse Gul~Atli, Samuel Marchal, and N.~Asokan.
\newblock {\em DAWN: Dynamic Adversarial Watermarking of Neural Networks.}
\newblock 2019.

\bibitem{ZBHRV16}
Chiyuan Zhang, Samy Bengio, Moritz Hardt, Benjamin Recht, and Oriol Vinyals.
\newblock Understanding deep learning requires rethinking generalization.
\newblock 2016.

\bibitem{CLEKS19}
Nicholas Carlini, Chang Liu, {\'U}lfar Erlingsson, Jernej Kos, and Dawn Song.
\newblock The secret sharer: Evaluating and testing unintended memorization in
  neural networks.
\newblock In {\em Proc. of {USENIX} Security}, pages 267--284. {USENIX}
  Association, 2019.

\bibitem{D06}
Cynthia Dwork.
\newblock Differential privacy.
\newblock In {\em Proc. of ICALP}, volume 4052 of {\em Lecture Notes in
  Computer Science}, pages 1--12. Springer, 2006.

\bibitem{HQH+19}
Huadi Zheng, Qingqing Ye, Haibo Hu, Chengfang Fang, and Jie Shi.
\newblock Bdpl: A boundary differentially private layer against machine
  learning model extraction attacks.
\newblock In {\em Proc. of ESORICS}, volume 11735 of {\em Lecture Notes in
  Computer Science}, pages 66--83. Springer, 2019.

\end{thebibliography}
